\documentclass[prb,superscriptaddress,floatfix,twocolumn,showpacs,amsfonts,amsmath,amssymb]{revtex4}
\usepackage{graphicx} 
\usepackage{dcolumn} 
\usepackage{bm} 
\usepackage{color}

\usepackage{tikz}

\begin{document}

\title{Measuring hole $g$-factor anisotropies using transverse magnetic focusing}

\author{Samuel Bladwell}
\affiliation{School of Physics, University of New South Wales, Sydney 2052, Australia}
\author{Oleg P. Sushkov}
\affiliation{School of Physics, University of New South Wales, Sydney 2052, Australia}

\begin{abstract}

Recent theoretical and experimental results from quasi-one dimensional 
heavy hole systems
 have suggested that heavy hole gases have a strongly anisotropic $g$ factor. 
In this theoretical paper, we propose a method for measuring this anisotropy
using transverse magnetic focusing (TMF). We demonstrate that for experimentally accessible fields, 
the $g$ factor anisotropy leads to a relative variation in the characteristic of spin-splitting of 
the TMF spectrum which allows for the measurement of the anisotropy of the $g$ factor. 
We show that this variation is insensitive to additional spin-orbit interactions,
and is resolvable with current devices. 

\end{abstract}

\maketitle

The strength of the coupling of an electron to a magnetic field in free space
 is defined by the Bohr magneton, $\mu_B$,
and the electron $g$ factor. 
Like free
electrons, quasi-particles in condensed matter systems couple to an applied magnetic 
field, however, the form and magnitude of the $g$ factor is strongly influenced by
the surrounding material. 
The ``renormalisation" of the $g$ factor can lead to effective 
$g$ factors for charge carriers in semiconductors orders of magnitude larger than 
the free space value. 
With such large values, an applied magnetic field can result in a significant 
change to transport properties, even in relatively weak magnetic fields.
This effect persists in reduced dimensional systems, and can be enhanced
or suppressed, and develop asymmetries depending on the confinement. In low dimensional 
hole systems
additional kinematic structures are possible, due to the holes angular momentum 
being $J=3/2$, with the Zeeman interaction in heavy hole systems depending on
both momentum and in-plane magnetic field\cite{Winkler2003, Li2016b}. Recent experimental
and theoretical results suggest a strongly anisotropic in-plane $g$-factor for two and 
quasi-one dimensional heavy holes systems\cite{Miserev2017a, Miserev2017b}. 

Measuring the $g$ factor anisotropy is difficult with typical transport techniques.
For instance, magnetic (Shubnikov de Haas) oscillations measure the total size of the Fermi
surface and therefore have no first order dependence on the anisotropy.  
Instead Shubnikov de Haas oscillations measure the total Fermi surface area. 
In this work we propose a method to measure the relative $g$ factors in 
hole systems, based around transverse magnetic focusing (TMF), which has a long history of
use in the measurement of the shape of the Fermi surface in both metals and 
semiconductors (see Fig. 1)\cite{Sharvin1965b, Sharvin1965a, Tsoi1974, Tsoi1999, Vanhouten1989}.
 When employed in systems with strong spin-orbit coupling, the first
magnetic focusing peak is spin split resulting in a ``double" focusing spectrum\cite{Rokhinson2004, Zulicke2007, Usaj2004, Reynoso2007, Schliemann2008, Kormanyos2010}.
 This spin-splitting can be directly translated to 
the strength of the spin-orbit interaction. We make use of this feature, combined
with a unique dependence on magnetic field rotations in TMF resulting from the anisotropy 
in the hole $g$ factor to determine the magnitude of the anisotropy. We then demonstrate that the effect is robust in the presence 
surface or bulk inversion asymmetry weaker that the Zeeman interaction due to the in-plane field, 
and provides a straightforward method of determining
$g_1$ and $g_2$.

\begin{figure}[h!]
    \begin{center}
     {\includegraphics[width=0.35\textwidth]{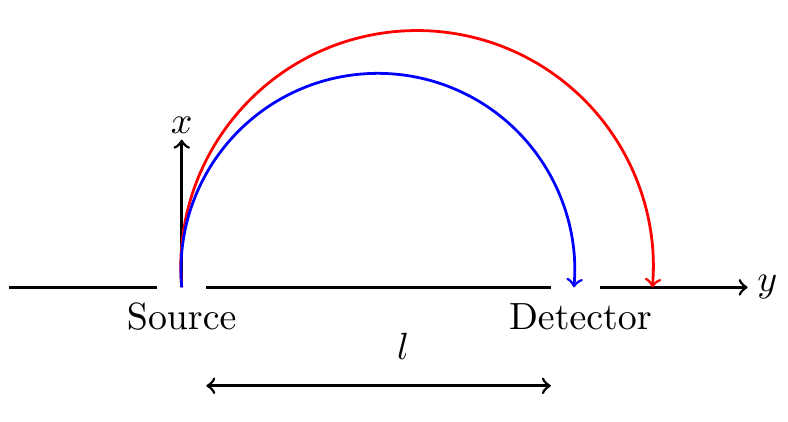}}
    \caption{Magnetic focusing setup, with focusing length, $l$. The source and 
    detector in hole systems are typically quantum point contacts (QPC). }
  \label{magfoc}    
  \end{center}
\end{figure}

We begin with kinematic structure leading to this $g$ factor anisotropy
in two dimensional  hole systems. Holes have a total
angular momentum of ${J} = 3/2$. At $k=0$, there are four degenerate states, 
typically denoted as ``light", $\pm1/2$,  and
``heavy", $\pm3/2$, holes due to the difference in effective mass. 
When confined to two dimensions only the heavy holes lie below the chemical 
potential. Coupling $J_z = 3/2$ to $J_z = -3/2$ requires $J_\pm^3$,
which is obtained with the combined action of the Luttinger term, $({\bf P \cdot J})^2$,\cite{Luttinger1955, Winkler2003}
and Zeeman interaction, ${\bf B \cdot J}$. Two kinematic structures are possible, 
$P_+^2 B_+ J_-^3$, or $P_+^4 B_- J_-^3 $. Since only the heavy holes lie below the chemical potential, 
it is convenient to work in the subspace $\pm3/2$, spanned by the Pauli matrices, with
$J_\pm^3 \rightarrow \sigma_\pm$. The kinematic structure is then
\begin{eqnarray}
{\cal H}_1 = \frac{g_1 \mu_B}{2} \left(B_+ p_+^2 \sigma_- + B_- p_-^2 \sigma_+\right)
\label{Zeeman1}
\end{eqnarray}
for the $g_1$ interaction and 
\begin{eqnarray}
\label{Zeeman2}
{\cal H}_2 = \frac{g_2 \mu_B}{2} \left(B_- p_+^4 \sigma_- + B_+ p_-^4 \sigma_+\right)
\end{eqnarray}
for the $g_2$ interaction. Here $p_\pm =  p_x \pm ip_y$ and $\sigma_\pm = \sigma_x + i \sigma_y$. 
Due to the momentum dependence of the interactions, the coefficients $g_1$ and $g_2$ are not 
dimensionless. For the following analytical calculations it is useful to consider the dimensionless
coefficients,
\begin{eqnarray}
\tilde g_1 = g_1 k_F^2 \quad \tilde g_2 = g_2 k_F^4
\label{dimensionless}
\end{eqnarray}
where $k_F = \sqrt{2m\varepsilon_F}$ is the Fermi momentum, and $\varepsilon_F$ is the Fermi energy.
Importantly, recent theoretical and
experimental work has shown that at experimental accessible densities, $\tilde g_2 $ can be comparable
to $\tilde g_1$\cite{Miserev2017a, Miserev2017b}.

\begin{figure}[t!]
     {\includegraphics[width=0.22\textwidth]{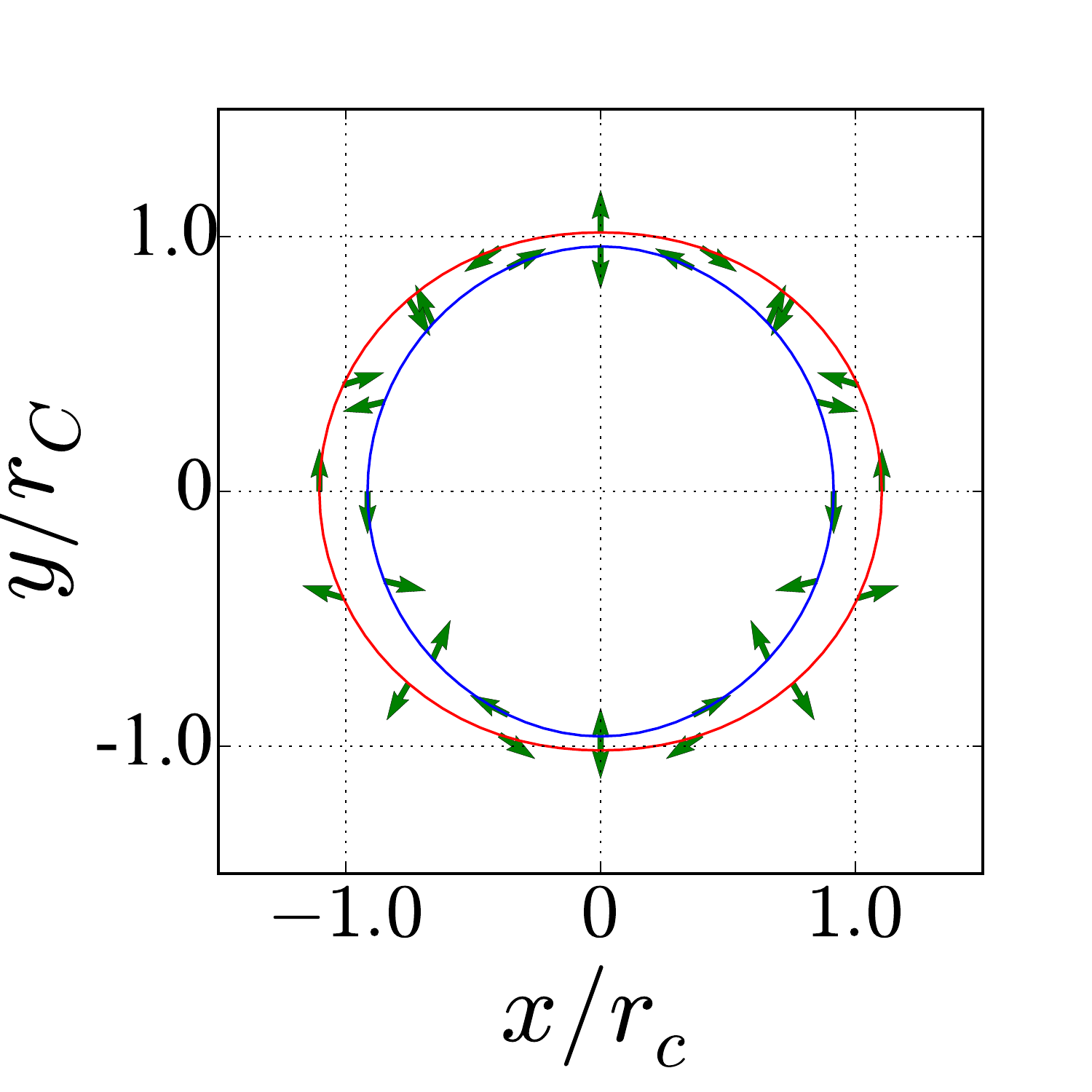}}
     {\includegraphics[width=0.22\textwidth]{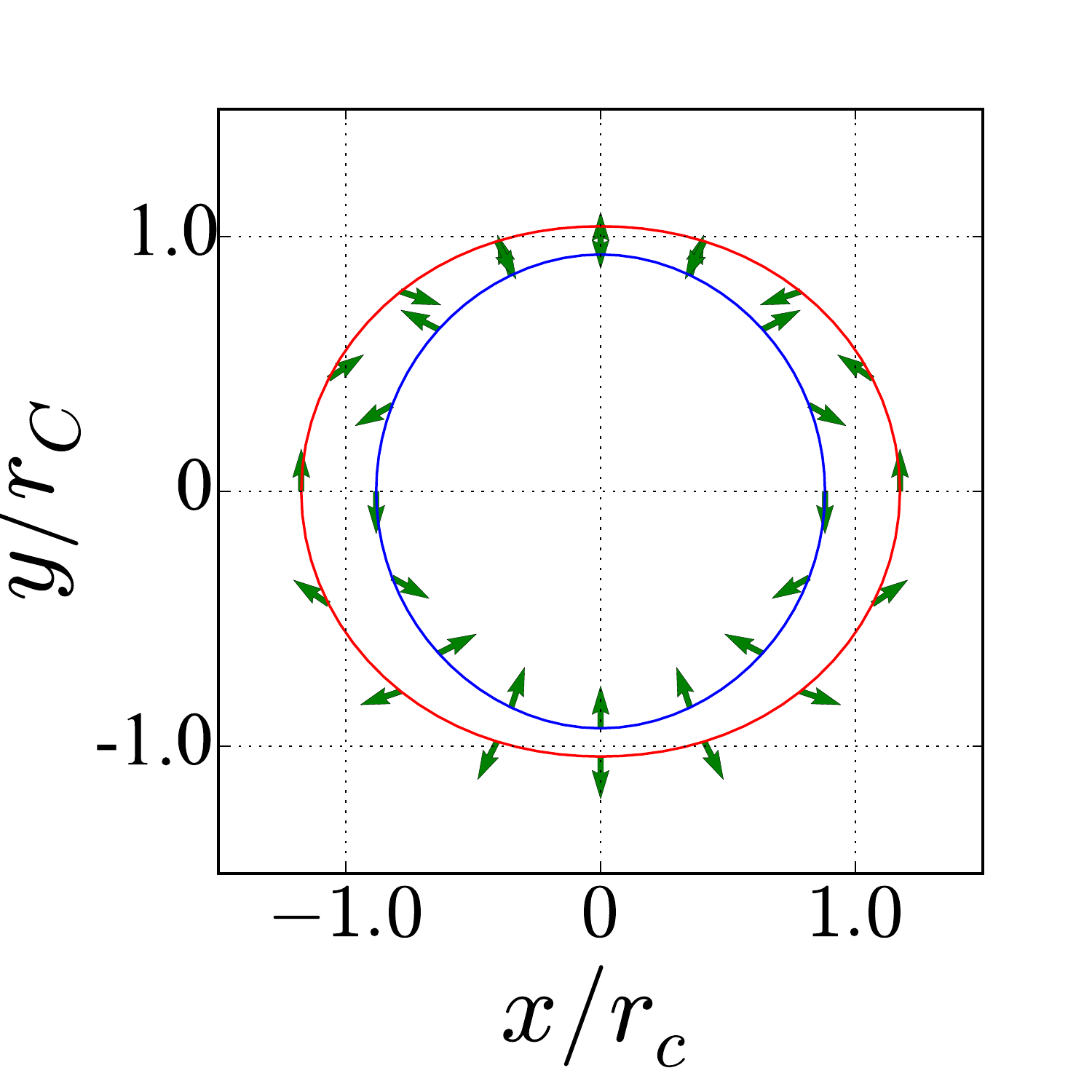}}
     	\caption{Cyclotron orbits for spin down (red) and up (blue) with the spin ($\left< {\boldsymbol{\sigma}} \right>$)
	orientation in the adiabatic limit. Left panel has $\tilde g_2 / \tilde g_1 = 0.5$, while 
	the right panel has $ \tilde g_1/  \tilde g_2 = 0.5$. We note that while the Fermi surfaces
	have a nearly identical shape, due to the different momentum dependence of the two 
	interactions, the spin dynamics are qualitatively different. }
\label{dispersion}
\end{figure}

The Hamiltonian for a hole system subject to these two respective Zeeman interactions
due to an in plane magnetic field, and some significantly weaker transverse focusing field is 
\begin{eqnarray}
{\cal H} &&= \frac{\hat{\boldsymbol{\pi}}^2}{2m} + {\cal H}_1 + {\cal H}_2 + \frac{g_z \mu_B}{2} B_z \sigma_z \\ \nonumber
&&= \frac{\hat{\boldsymbol{\pi}}^2}{2m} + {\cal B} \left( \hat{\boldsymbol{\pi}} \right) \cdot \boldsymbol{\sigma} \\ \nonumber
&&\hat{\boldsymbol{\pi}} = \hat{\bm p} - e {\bm A} \nonumber
\label{ham2}
\end{eqnarray}
where ${\bm A}$ is the vector potential. The equations of motion are
\begin{eqnarray}
\dot{\hat{\boldsymbol{\pi}}} = i \left[ {\cal H}, \hat{{\boldsymbol{\pi}}} \right] = m\omega_c \hat{\bf v} \times {\bf n}  \\ \nonumber
\dot{\boldsymbol{\sigma}} =  i \left[ {\cal H}, {{\boldsymbol{\sigma}}} \right] = - {\cal B} \left( \hat{\boldsymbol{\pi}} \right) \times \boldsymbol{\sigma}
\label{eqmotion}
\end{eqnarray}
where $\omega_c = e B_z/m$.
Solving these equations of motion requires some approximation of 
spin-dynamics. 
In the case of TMF, the appropriate approximation for the resolution of 
a double focusing peak is adiabatic spin dynamics, where the spin follows the instantaneous 
effective magnetic field, 
$\left< {\boldsymbol{\sigma}}\right> =s{\cal B}/{|{\cal B}|}$, where $s=\pm1$ is a pseudo scalar defining
the spin projection\cite{Zulicke2007}.
The resulting semi-classical equation is obtained using the method of  Ref. \cite{Bladwell2015, Bladwell2017e}, 
\begin{eqnarray}
{\bf r}(\theta(t)) = \frac{\boldsymbol{\pi}(\theta(t)) \times {\bf n}}{eB_z} - \frac{\boldsymbol{\pi}(\theta(0)) \times {\bf n}}{eB_z}
\label{traj}
\end{eqnarray}
where the momentum $\boldsymbol{\pi}$ depends on the spin-state of the hole, and $\theta(t)$ is the 
polar (running) angle and is a function of time. 
Physically, this corresponds
to the classical cyclotron motion of a quasi particles with different cyclotron radii depending on the 
spin projection. We present both
cyclotron orbits and spin orientations in Fig. \ref{dispersion}. 

\begin{figure}[t!]
\begin{center}
     {\includegraphics[width=0.45\textwidth]{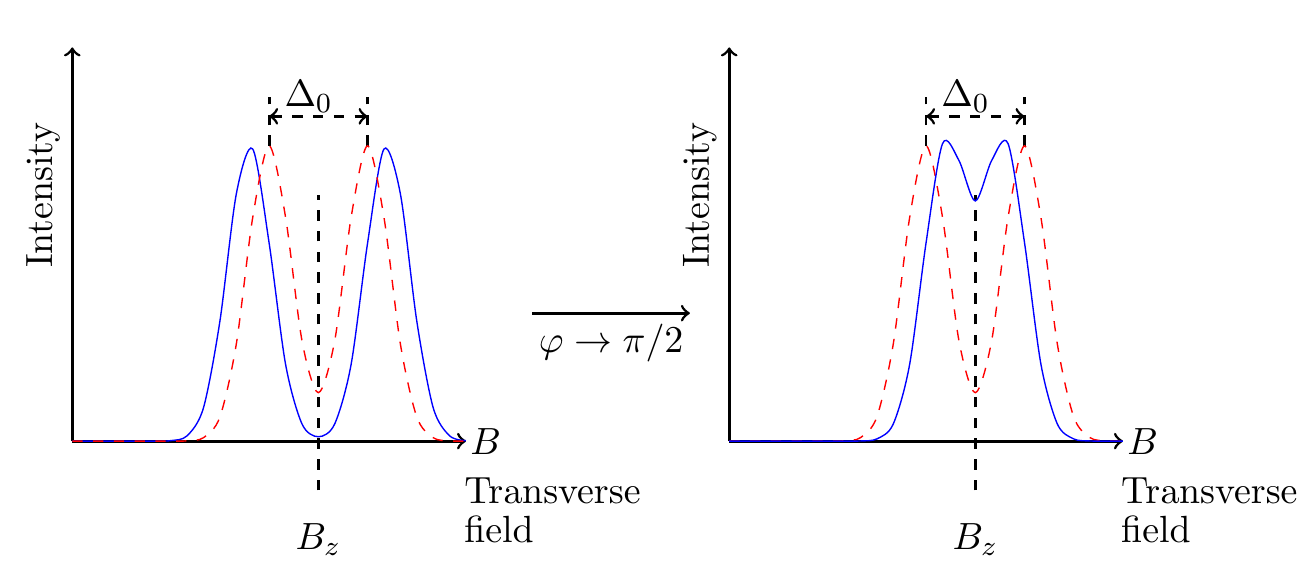}}
\caption{Cartoon of the variation in the focusing splitting. Red dashed curve is the TMF spectrum without any
$g$ factor asymmetry, with splitting magnitude $\Delta_0$, with ${\Delta_0} =  \tilde g_1 \mu_B B_{||}/\varepsilon_F$. 
The blue curve is the TMF spectrum, including the $g$ factor anisotropy, 
with $\varphi=0$ for the left panel, and $\varphi = \pi/2$ for the 
right panel. The focusing field without spin splitting, $B_z = 2k_F/eL$, is marked with a dashed black line.}
\label{cartoon}
\end{center}
\end{figure}

To explore the dynamics analytically, we consider the following approximation for the 
spin split momentum, $\boldsymbol{\pi}$, 
\begin{eqnarray}
\boldsymbol{\pi} = \hbar k_{F, s} (\cos \theta(t), \sin \theta(t), 0) \\
k_{F, s}  = k_F \left(1 + \frac{s|{\cal B}|}{2 \varepsilon_F} \right) \nonumber
\label{approxk}
\end{eqnarray}
where the total effective magnetic field is
\begin{eqnarray}
|{\cal B}| = \mu_B B_{||} \sqrt{ \tilde g_1^2 + \tilde g_2^2 + 2 \tilde g_1 \tilde g_2  \cos(2\theta - 2\varphi) } \nonumber \\
\label{totalfield}
\end{eqnarray}
with an in-plane field angle $\varphi$, and
\begin{eqnarray}
{\bm B} = \left( B_{||} \cos{\varphi}, {B_{||}} \sin\varphi, B_z \right)
\end{eqnarray}
being the magnetic field applied to the sample. To satisfy the requirement for adiabatic spin dynamics, 
we need to ensure that the magnetic field remains sufficiently large. If $\tilde g_1 > \tilde g_2$ 
the approximate condition of adiabatic spin dynamics is
\begin{eqnarray}
|{\cal B}| \gg \frac{1}{|{\cal B}|}\frac{\partial {\cal B}}{\partial t} \sim 2 \omega_c
\end{eqnarray} 
where the factor of $2$ comes for the two rotations of the spin-orbit field for each rotation in momentum space,
see Fig. \ref{dispersion}.
 In this adiabatic regime, 
the classical focusing peak where interference effects are neglected corresponds 
to an injection angle of $\theta=0$\cite{Bladwell2017}. 
The focusing 
length, l, from Eqs. \eqref{traj} and \eqref{approxk} is
\begin{eqnarray}
l = y(\theta = \pi) \approx \frac{\hbar k_{F, s}(\theta=0) + \hbar k_{F,s} (\theta = \pi)}{e B_z}
\label{foc1}
\end{eqnarray}
which is analogous to the case of classical TMF in metals \cite{Sharvin1965a}. 
We have
cast the above result in terms of spatial variation of the peaks; 
in TMF the detector and collector are  
fixed, and the focusing field, $B_z$ is varied instead, 
\begin{eqnarray}
B_z  \approx \frac{\hbar k_{F, s}(\theta=0) + \hbar k_{F,s} (\theta = \pi)}{e L}
\label{foc1}
\end{eqnarray}
For $2\omega_c \ll |{\cal B}| \ll \varepsilon_F$, using 
Eqs. \eqref{foc1} and \eqref{approxk}, 
\begin{eqnarray}
\frac{\delta B_z}{B_z} \approx \frac{k_+(\varphi) - k_-(\varphi)}{k_F}
\label{effectivesplitting}
\end{eqnarray}
where $\delta B_z$ denotes the splitting between the spin-split focusing peaks. Fig. \ref{magfoc} corresponds
to a fixed $B_z$ and a varied focusing length $l$; in a real experiment $l$ is fixed and $B_z$ is varied, see Fig. \ref{cartoon}.

\begin{figure}[t!]
\begin{center}
     {\includegraphics[width=0.45\textwidth]{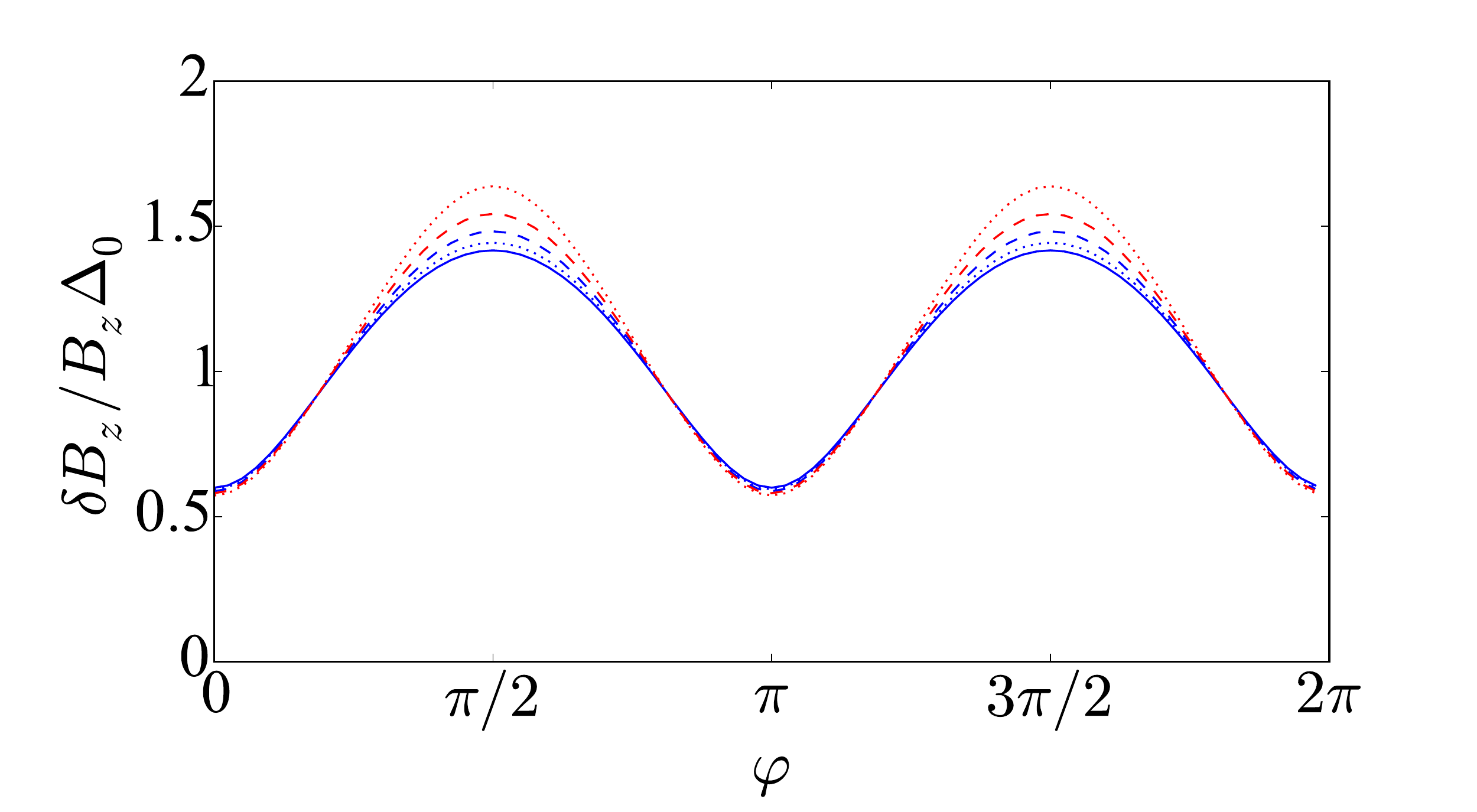}}
      \caption{
      The field angle dependence of the first focusing peak splitting, with increasing 
	in plane magnetic field, from 1T to 3.5T, with $\tilde g_2 = 1$ and $\tilde g_1 = - 2.5$.
	The deviation from Eq. \eqref{g1ong2} is the 
	result of the non parabolic terms in the dispersion. \newline 
	}
\label{field angle}
\end{center}
\end{figure}

We are now in a position to explore the angular dependence of in-plane field response 
in the TMF spectrum. We consider the case of a relatively short focusing length, of 
$l =1000$nm, with a hole density $n = 1.65 \times 10^{11}$cm$^{-2}$. 
Let us start with the case where all other spin-orbit interactions 
have been tuned to be small. For quantum wells grown along high symmetry 
axes, this is a reasonable approximation. For the 
case where $\tilde g_1 \gg \tilde g_2$, by expanding in terms of $\tilde g_2/\tilde g_1$ 
we obtain the following approximate 
analytical expression for the angular dependence, 
\begin{eqnarray}
\frac{\delta B_z}{B_z \Delta_0} \approx \left(1 + \frac{\tilde g_2}{\tilde g_1}\cos(2\varphi)\right)
\label{g1ong2}
\end{eqnarray}
where we have introduced 
the dimensionless splitting, ${\Delta_0} = \tilde g_1 \mu_B B_{||}/\varepsilon_F$. 
In Fig. \ref{field angle} we plot the fractional focusing field splitting, $\delta B_z/B_z\Delta_0$, as a function of the 
in plane field angle, $\varphi$. Here $\tilde g_1 = -2.5$ and $\tilde g_2 = 1$. 
For GaAs quantum wells, $|\tilde g_1|, |\tilde g_2| < 3$, dependent on the 
manner of the confinement\cite{Miserev2017b}. 
 
In practice other spin-orbit
 interactions due to bulk and surface inversion symmetry may 
 not be small. To examine the influence of additional spin-orbit interactions
 we consider a Rashba spin-orbit interaction, ${\cal H}_R = i\gamma_R p_+^3 \sigma_-/2 + h.c.$, resulting
 from an asymmetric confining potential. The Rashba induces a 
 spin-splitting in the hole gas of $\Delta_{R} = \gamma_3 \hbar^3 k_F^3/\varepsilon_F$.  
 We include the Rashba term in the Hamiltonian, 
 Eq. \eqref{ham2}, and use the aforementioned method to determine the variation in
 the focusing field.  In Fig. \ref{field angle} we present the 
response to in-plane magnetic field rotations, with varying strength of the Rashba spin-orbit interaction. 
Provided
$\Delta_R < \tilde g_1 \mu_B B_{||}/\varepsilon_F$ and $\Delta_R < \tilde g_2 \mu_B B_{||}/\varepsilon_F$, 
there is minimal variation in the 
magnetic focusing field splitting, $\delta B_z$.
In general, 
spin-orbit interactions which are odd in momentum such as the Dresselhaus and Rashba interactions
will only weakly perturb the variation in the TMF peak spacing. 

\begin{figure}[t!]
\begin{center}
     {\includegraphics[width=0.45\textwidth]{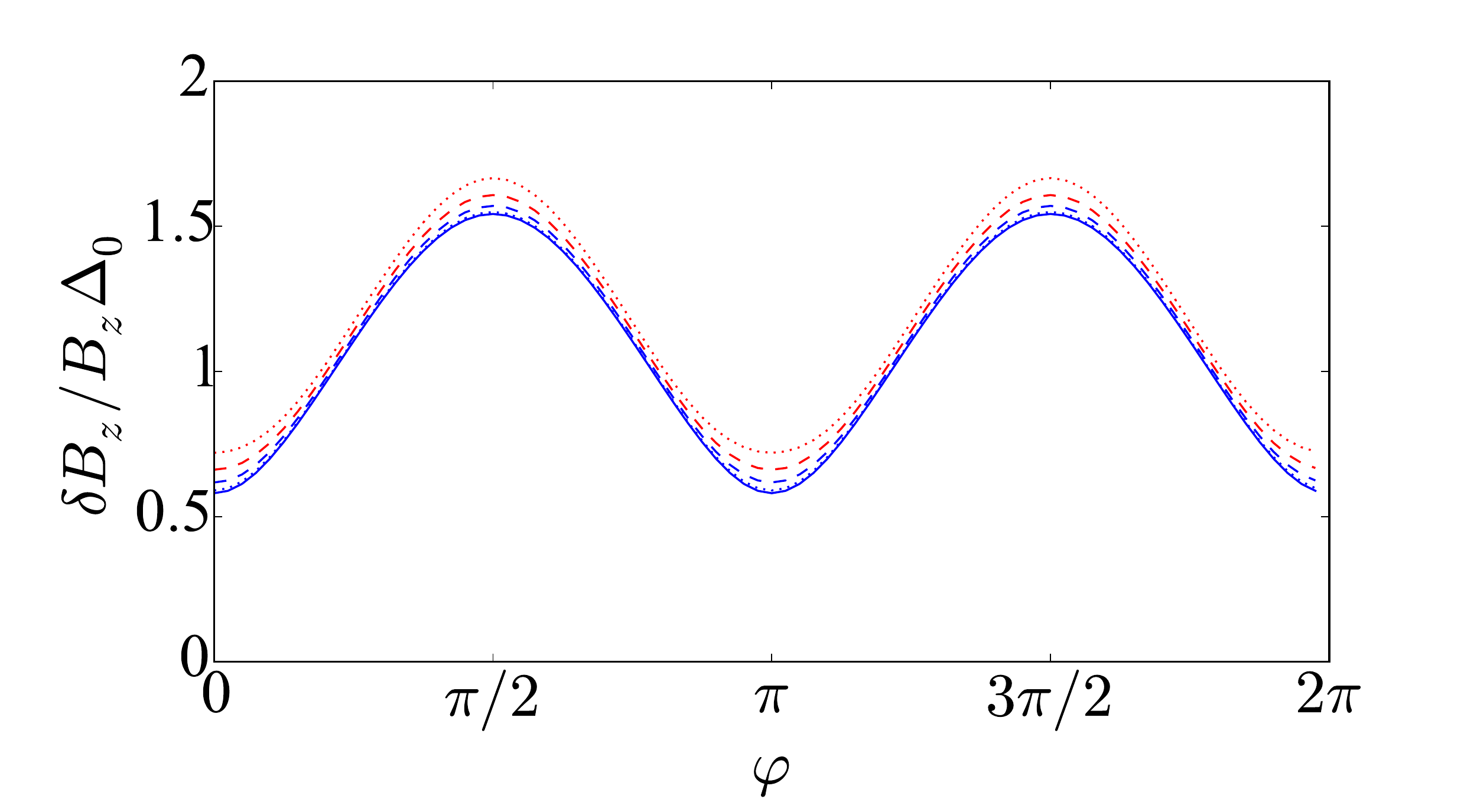}}
     	\caption{Relative focusing field splitting as a function of the 
     	in-plane magnetic field angle, $\varphi$, with $B_{||} = 2.5$T, $\tilde g_1 = -2.5$, $\tilde g_2 = 1$. Here we vary the
	strength of the Rashba interaction,  $\Delta_{R}$, in the range
	 $0 <\Delta_R<0.1\epsilon_F$. The maximum value is $\Delta_{R} \sim 0.2$meV.}
\label{field angle 2}
\end{center}
\end{figure}

The measurement of $\tilde g_2/ \tilde g_1$ depends only on the classical focusing field, 
and is therefore independent of the source and detector. However,  the effect must be larger that the 
spread of the focusing peaks to be observed, $\delta B_z/B_z \ge B_{FWHM}/B_z$, where
$B_{FWHM}$ is the full width half maximum of the focusing peak. 
Recent hole experiments with spin-splitting induced via a large Rashba type interaction
have $B_{FWHM} \sim 0.02$T\cite{Rokhinson2004}, giving a ratio of $B_{FWHM} /B_z \sim 0.1$. 
To make a direct comparison between this and our results, we consider an in-plane field of $B_{||} = 3.5$T 
and $\tilde g_1 = -2.5$. 
with $\Delta_0 \approx 0.2$. Comparing to Fig. \ref{field angle}, the minimum value at $\theta = \pi$, 
corresponds to an effective splitting $\delta B_z/B_z > 0.1$. Hence the two peaks are resolvable over the full range 
of $\varphi$ at $B_{||} = 3.5$T. 

Finally we turn our attention to the assumption of adiabatic spin dynamics that
we have employed in preceding calculations.
As has been noted, $2 \omega_c \ll |{\cal B}_{min}|$.
The minimum value of ${\cal B}$, 
\begin{eqnarray}
|{\cal B}_{min}| = \left(|\tilde g_1| - |\tilde g_2|\right) \mu_B B_{||} 
\label{zeemanmin}
\end{eqnarray}
which results in the following condition, 
\begin{eqnarray}
4 \frac{m}{m^{*}}\frac{B_z}{B_{||}} < |\tilde g_1| - |\tilde g_2| \approx 1
\label{adiabat1}
\end{eqnarray}
The fraction $B_z/B_{||}$ is the ratio between the in-plane field, and the focusing
field, while $m^*$ is the effective mass.  
For a typical device, $B_z\sim 0.1$T, while $B_{||}$ can be several Tesla, and 
in quantum wells, $m^*\sim 0.2m$. 
We can compare this to some recent results in GaAs heavy
hole quantum wells. The commensurate
criterion to \eqref{adiabat1} is
\begin{eqnarray}
3 \omega_c < \Delta_R
\label{adiabat2}
\end{eqnarray}
where $\Delta_R$ is the strength of the Rashba splitting, $\Delta_R \sim 0.2 \varepsilon_F$. 
Which can be converted to an expression in terms of the Fermi momentum, $k_F$ and focusing
length $L$, 
\begin{eqnarray}
\frac{12}{k_F L } < \Delta_R
\label{adiabat3}
\end{eqnarray}
For this Rashba hole system, $k_F L \sim 100$, so $\Delta_R k_F L /12 \sim 1/2$, 
while the two spin-split peaks are still clearly observable. 
Comparing this to Eq. \eqref{adiabat1}, $B_{||} \sim 4$T is sufficient to satisfy this condition. 
Taken together with considerations of the spread of the focusing peak, 
we can conclude that it is possible to measure anisotropies in the in-plane $g$
factor in hole systems using TMF. 

In summary, we have shown TMF can be used to determine the relative magnitude of 
$\tilde g_1$ and $\tilde g_2$, via the unique dependence on the in plane magnetic field angle, $\varphi$. Futhermore, 
this dependence is robust with the addition of residual spin-orbit interactions. 
Based on 
results from TMF in heavy hole gases, variation the focusing field is 
significantly larger than the broadening 
due to both scattering and the finite size of injectors and detectors at experimentally accessible in plane fields.
This work was supported by the
Australian Research Council Centre of Excellence in Future Low-Energy Electronics Technologies
 (project number CE170100039) and funded by the Australian Government.
 The authors would like to thank Dima Miserev for his valuable discussions.

\bibliography{bibliography}

\end{document}